\documentclass[aps,pre,preprint,superscriptaddress]{revtex4}

\usepackage{amsmath}
\usepackage{graphics}

\newcommand{\DeltaG}{\Delta G}
\newcommand{\DeltaMu}{\Delta \mu}
\newcommand{\Deltagamma}{\Delta \gamma}
\newcommand{\DeltaCp}{\Delta C_\mathrm{p}}
\newcommand{\gammaOW}{\gamma_{\mathrm{ow}}}          
\newcommand{\gammaWV}{\gamma}          
\newcommand{\nstar}{n^*}
\newcommand{\cmc}{\rho_{\mathrm{cmc}}}

\newcommand{\muS}{\mu_{\mathrm{S}}}

\begin{document}

\title{Micelle Formation and the Hydrophobic Effect}
\author{Lutz Maibaum}
\affiliation{Department of Chemistry, University of California, Berkeley, CA 94720}
\author{Aaron R. Dinner}
\affiliation{Department of Chemistry, University of Chicago, Chicago, IL 60637}
\author{David Chandler\footnote{Corresponding author.}}
\affiliation{Department of Chemistry, University of California, Berkeley, CA 94720}

\date{January 14, 2004}

\begin{abstract}
The tendency of amphiphilic molecules to form micelles in aqueous solution
is a consequence of the hydrophobic effect. 
The fundamental difference between micelle assembly and macroscopic phase
separation is the stoichiometric constraint that frustrates the
demixing of polar and hydrophobic groups. 
We present a theory for micelle assembly that combines the account of
this constraint with a description of the hydrophobic driving
force. The latter arises from the length scale dependence of aqueous
solvation. The theoretical predictions for temperature dependence and
surfactant chain length dependence of critical micelle
concentrations for nonionic surfactants agree favorably with
experiment.
\end{abstract}

\maketitle

\section{Introduction}

This paper concerns the formation of micelles, which are the simplest
form of amphiphilic assemblies. Our treatment of this phenomenon is based
upon the length scale dependence of hydrophobic effects~\cite{LCW, DC}. Namely,
the free energy to solvate small hydrophobic molecules scales linearly
with solute volume, while that to solvate large hydrophobic species
scales linearly with surface area. The crossover from one regime to
the other occurs when the oily species presents a surface in water
extending over about $1 \, \mathrm{nm}^2$. Due to these
contrasting scalings, the free energy to solvate a collection of small
oily species that are well separated in water can exceed that of solvating a
large cluster of these same species. The resulting free energy difference is the
hydrophobic driving force for assembly. While this force that derives
from this scaling difference is surely
the mechanism that drives oil-water phase separation, one may wonder if it is
also applicable in the case of amphiphilic assembly where interfaces between
solute clusters and water contain polar or charged head-groups as well as
oil. In this paper, we argue that it is indeed applicable.

The primary difference between oil-water phase separation and amphiphilic
assembly is due to stoichiometry. Each amphiphilic molecule contains an oily
species that is constrained to remain within a molecular length of a
hydrophilic species. For large clusters of amphiphiles, this constraint
leads to an entropic cost for clustering oily components that grows faster
than cluster volume. Thus, unlike simple oil-water phase separation which forms
macroscopic domains, growth of amphiphilic assemblies is limited to
mesoscopic domains. Our treatment of this effect uses an estimate of entropy~\cite
{Woo96} obtained from an electrostatic analogy for stoichiometric
constraints~\cite{DeGennes, Stillinger, Wu92}. In the next section, we
describe this estimate along with the other factors that contribute to
micelle formation. The resulting expression for the critical micelle
concentration, $\cmc$, is then compared with experiments in Section III. In
particular, we show that our expression yields good predictions of this
concentration, changing with temperature and with amphiphile chain length in
accord with experimental observations. An appendix is used to augment the
discussion in Section II.

\section{Theory}

\subsection{Law of mass action}

We consider an aqueous solution of neutral amphiphilic molecules
(i.e., non-ionic surfactants), each of which has
a single alkyl chain as its hydrophobic tail. In general, amphiphiles can
form aggregates of various sizes and shapes. We will assume each micelle is
spherical and neglect the effects of fluctuations in micelle size and
shape. Thus, we imagine that each surfactant molecule exists either as a
monomer or as part of a spherical $n$-mer. We denote the number densities
of the monomers and $n$-mers by $\rho _{1}$ and $\rho _{n}$, respectively,
so that the total surfactant concentration is given by $\rho =\rho
_{1}+n\rho _{n}$.

\begin{figure}[t]
\resizebox{\columnwidth}{!}{
\includegraphics{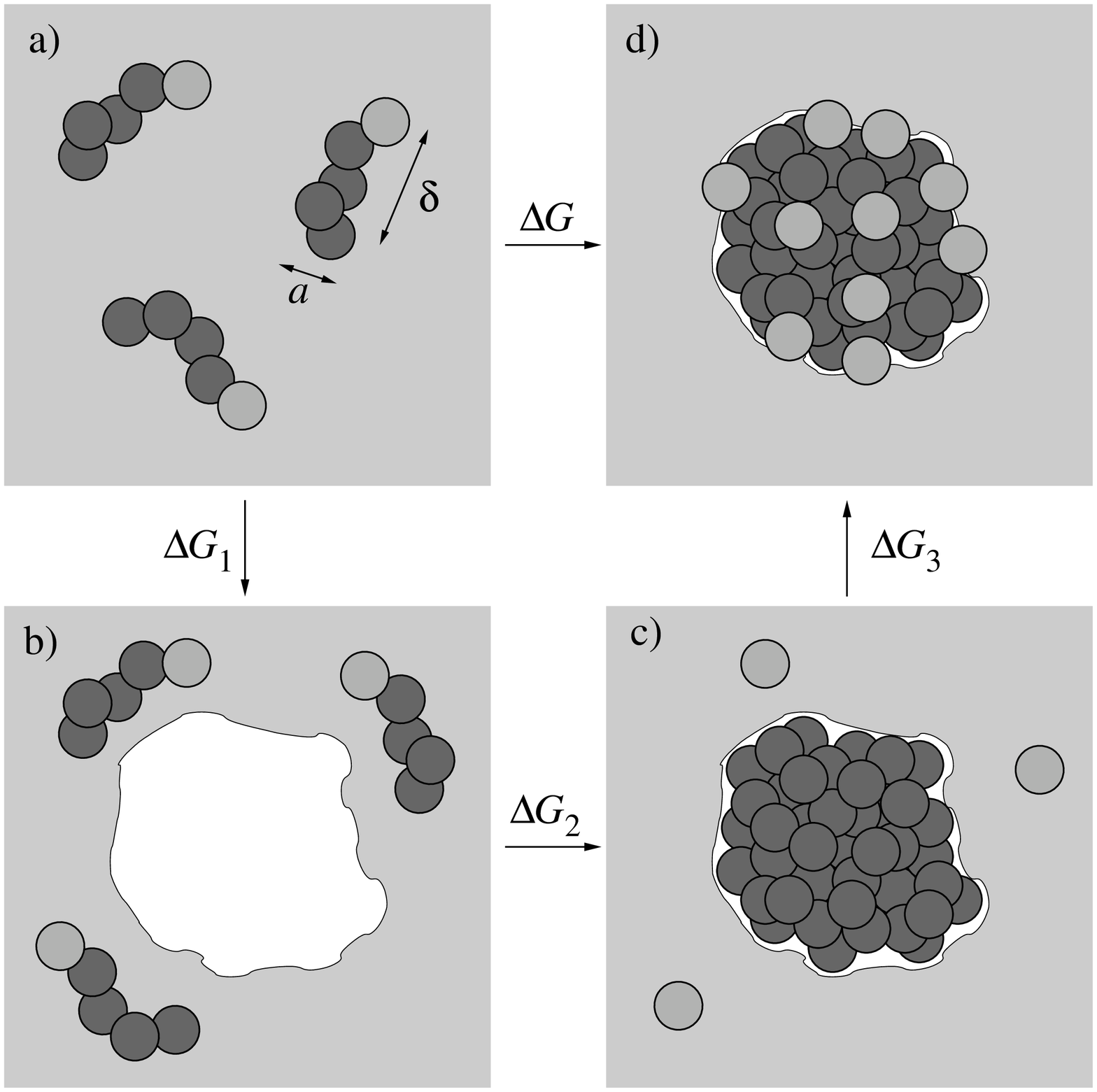}
}
\caption{\label{fig:micelles}
Thermodynamic cycle of micelle formation: the process of assembling $n$
separated amphiphiles (a) to a micelle (d) can be
performed in three steps: \textit{1)} Creating a cavity in the solvent
(light gray) (b); \textit{2)} Transferring the hydrophobic chains
(dark gray) from the aqueous solution
into the cavity (c); \textit{3)} Distributing the polar units (gray)  over the
surface of the cavity, and reconnecting them to the hydrophobic groups
(d).
}
\end{figure}

The concentrations of monomers and micelles are related by the law of mass
action \cite{greenbook},
\begin{equation}
\rho _{n}a^{3}=\left( \rho _{1}a^{3}\right) ^{n}\exp \left( -\beta
\DeltaG \right) ,  \label{eq:MassAction}
\end{equation}
where $\beta $ denotes inverse temperature (i.e., $\beta^{-1} =
k_{\mathrm{B}}T)$, $a$ is a microscopic length that specifies the
standard state convention, and
\begin{equation}
\DeltaG=f_{n}-nf_{1}
\end{equation}
is the driving force for assembly, namely, the free energy of the $n$-mer, $%
f_{n}$, relative to that of $n$ monomers, $nf_{1}$. We
take $a$ to be approximately the girth of a surfactant molecule, see
Fig. \ref{fig:micelles}.

For large $n$, eq~\ref{eq:MassAction} implies the existence of a threshold
concentration of surfactant molecules $\cmc$, at which the
density of aggregates becomes significant. Because this crossover is
precipitous, its location is almost independent
of the specific definition of the threshold as long as it is
physically sensible. Specifically, to within corrections of order $n^{-1}\ln n$,
\begin{equation}
\ln \cmc a^{3}=\beta \DeltaG / n^{*} .
\label{eq:cmcgeneral}
\end{equation}
The driving force per surfactant, $\DeltaG / n$, is a function of $n$, and it
is to be evaluated at the most probable aggregation number, $n^{*}$. This
number is the value of $n$ that minimizes $\DeltaG / n$. Equation
\ref{eq:cmcgeneral} is discussed further in the Appendix.

\subsection{Driving force}

The contributions to $\DeltaG$ can be found in three steps, employing
the thermodynamic cycle illustrated in Fig.~\ref{fig:micelles}.

1) \textit{Creation of a cavity.} A micelle will fill a region vacated by
water. Assuming the extent of the surface is at least 1 nm$^{2},$ the free
energy to create this cavity is
\begin{equation}
\DeltaG_{1}=\gammaWV \, A ,  
\label{eq:delta-G_1}
\end{equation}
where $A$ denotes the surface area of the cavity, and $\gammaWV$ is the
water--vapor surface tension. In general, there is also pressure--volume
work for forming a cavity in a liquid. For water at standard
conditions, pressure is sufficiently small that this contribution is
negligible for cavities with diameters less than 5 nm. We will limit
our consideration to sizes within this range.

2) \textit{Filling the hydrophobic core}. Imagine disconnecting
each hydrophobic tail in a surfactant from its respective hydrophilic head
group and moving the hydrophobic tail from water into the micelle core (we
will reconnect the heads and tails in Step 3). A total of $n$ tails must be
moved to fill the cavity formed in Step~1. As such, one part of the free
energy to fill the cavity is $-n\,\DeltaMu $, where $-\DeltaMu $ is the
free energy change in transferring the hydrophobic tail (e.g., an alkane
chain) from water into the oily hydrophobic core. An additional part of the
free energy for filling the cavity is an interfacial contribution due to the
presence of van der Waals attractions between oil and water. These
interactions cause the oil-water surface tension, $\gammaOW$, to be
lower than the water-vapor surface tension, $\gammaWV$ (see, for
instance, Ref.~\onlinecite{Huang-DC02}). Thus, the free energy for
filling the cavity is
\begin{equation}
\DeltaG_{2}=-n\,\DeltaMu -\Deltagamma \, A ,  
\label{eq:delta-G_2} 
\end{equation}
where $\Deltagamma =\gamma -\gammaOW$.

The interior of a micelle is densely packed and much like a hydrocarbon
liquid~\cite{Tanford}. Thus, $\DeltaMu $ is close to the transfer free
energy for moving the associated alkane chain from oil into
water. However, it is slightly smaller than this value because the
environment of an alkane chain in a micelle interior is more
confining than that in bulk oil~\cite{Tanford}. The numerical
consequence of this small difference will be discussed in Sec.~III. To
the extent that the micelle is spherical, $A=4\pi L^{2}$, where $L$ is
the micelle radius. Since the interior is densely packed, $L$ is given
by $4\pi L^{3}/3=n\delta a^{2}$, where $\delta $ is the mean length
over which a polar head group is separated from an alkyl group within
a surfactant molecule, see Fig. 1. From these considerations,
\begin{equation}
\DeltaG_{1}+\DeltaG_{2}=-n\,\DeltaMu +g\,n^{2/3},
\label{eq:delta-G_1+delta-G_2}
\end{equation}
where $g=(36\pi )^{1/3}(\gammaOW\,a^{2})(\delta
/a)^{2/3}\approx 4.8\times (\gammaOW\,a^{2})(\delta
/a)^{2/3}$.

The right-hand side of eq \ref{eq:delta-G_1+delta-G_2} is essentially the
free energy for nucleating oil clusters in water~\cite{Barrat-Hansen}. It is
the hydrophobic driving force identified in the Lum-Chandler-Weeks theory~%
\cite{LCW}. The first term is proportional to the volume of hydrophobic units.
The second term is proportional to the area of the interface. The first term is
extensive in $n$ and dominates at large $n$. Thus, if only $\DeltaG_{1}$
and $\DeltaG_{2}$ were significant, the strength of the driving force
would grow without bound leading to macroscopic clusters. However, these
contributions are counter balanced by a third term that we consider now.

3) \textit{Placing hydrophilic head groups on micelle surface}. In the
final step, the hydrophilic head groups are reconnected to the hydrophobic
tails, placing them at the water--oil interface so as to maintain a favorable
solvation energy. This positioning is to be done while simultaneously
enforcing the connectivity between heads and tails and while also maintaining the
densely packed interior. These conditions result in an entropic cost that
increases super-extensively with aggregate size. The form of this third
contribution to the driving force is conveniently estimated from the
electrostatic analogy of stoichiometric constraints~\cite{DeGennes,
Stillinger}. The result is~\cite{Woo96}
\begin{equation}
\DeltaG_{3}=h \, n^{5/3} / \beta ,  
\label{eq:delta-G_3}
\end{equation}
where $h=(3/(4\pi ))^{2/3}(96/49)(a/\delta )^{4/3}\approx 0.75\times %
(a/\delta )^{4/3}$. In employing this analogy, it is important to note
that the micelle volume is essentially that of the densely packed
alkyl chains.

\subsection{Micelle size and critical micelle concentration}

Combining the three contributions discussed above gives the driving
force in units of $k_{\mathrm B} T$:
\begin{equation}
\beta \DeltaG \approx -n\,\beta \,\DeltaMu +\beta
\,g\,n^{2/3}+h\,n^{5/3}.  \label{eq:DeltaG}
\end{equation}
Minimization of $\DeltaG / n$ therefore gives
\begin{equation}
\nstar \approx \beta g/2h=\left( 49\pi /48\right) \,\beta \,\gamma \,\delta
^{2}.  \label{eq:n-star}
\end{equation}
With this aggregation number, eqs \ref{eq:cmcgeneral} and \ref{eq:DeltaG} yield
\begin{equation}
\ln \cmc a^{3}=c\,\left( \beta \gammaOW a^{2}\right)
^{2/3}\,-\,\beta \,\DeltaMu ,  
\label{eq: result}
\end{equation}
where $c=(5832/49)^{1/3}\approx 4.9$.

\section{Comparison With Experiment}

Equation \ref{eq: result} is the principal result of this paper. It
expresses the critical micelle concentration in terms of measured quantities
and one adjustable parameter, the molecular length scale $a$. As such, its
validity is easily checked. Here we do so by considering $m$-alkyl
hexaoxyethylene glycol monoethers, C$_{m}$E$_{6}$, i.e., ($\mathrm{CH}_{3})(%
\mathrm{CH}_{2})_{\mathrm{m-1}}(\mathrm{OCH}_{2}\mathrm{CH}_{2})_{6}\mathrm{%
OH}$. The critical micelle concentrations for this class of nonionic
surfactants have been determined experimentally.

The surface tension is required to compare eq \ref{eq: result}
with experimental results. At room temperature, the oil--water surface
tension is $\gammaOW \approx 51\,\mathrm{mN}/\mathrm{m}$~\cite
{oilwatersurfacetension}. Were we to neglect $\Deltagamma ,$ we would
instead take the room temperature value of the water--vapor surface tension, $\gammaWV =72\,\mathrm{mN}/\mathrm{m}$ \cite{CRC}. The
difference $\Deltagamma =\gamma -\gammaOW$ is mainly
enthalpic~\cite{Huang-DC02}. As such, a good approximation for the
temperature dependence of the required surface tension is $d\gammaOW/dT\approx d\gammaWV /dT\approx -0.17\,\mathrm{mN}/\mathrm{m}/\mathrm{K}$~
\cite{CRC}.

The transfer free energy, $\DeltaMu$, is also needed. To the extent that
the micelle interior is like a bulk hydrocarbon liquid, it can be obtained from
solubility measurements \cite{McAuliffe66, McAuliffe69}. This free energy
change for transferring alkane chains with fewer than 12 carbons from oil to
water depends linearly on the number of carbons, $m$. The linear fit to
experimental data is $\DeltaMu_{0}(m)=(2.25+0.9\,m)\,\mathrm{kcal}/%
\mathrm{mol}$ at room temperature. We have used the subscript ``0'' to
indicate that this free energy for transferring between bulk phases must
differ to some extent from that for transferring from a micelle interior to
water. Indeed, experimental evidence for alkane chains with $m<6$ \cite
{Tanford} indicates that $\DeltaMu (m) \approx (1.9+0.77m)
\mathrm{kcal} / \mathrm{mol} \equiv \DeltaMu_{0} (m) +\Delta \DeltaMu (m)$.

The temperature dependence of $\DeltaMu $ can be estimated from the
observation of convergence temperatures for transfer entropies and
enthalpies, $T_{S}^{*}$ and $T_{H}^{*}$, respectively~\cite{Murphy90,
Murphy94}. Namely,
\begin{equation}
\DeltaMu (T,m) \approx \DeltaCp (m)\left[ (T-T_{H}^{*})-T\ln
(T/T_{S}^{*})\right] ,  \label{eq:delta-mu}
\end{equation}
where $T_{S}^{*}=112\,\mathrm{C}$ and $T_{H}^{*}=22\,\mathrm{C}$ \cite
{Baldwin86}.  The heat capacity $\DeltaCp \left( m\right) $ is chosen so as to fit the
experimental $\DeltaMu (m)$ at room temperature, i,e.,
\begin{equation}
\DeltaCp (m)=\left[ \DeltaMu_{0}(m)+\Delta \DeltaMu (m)\right] /\left[
(298\mathrm{K}-T_{H}^{*})- 298\mathrm{K} \ln (298\mathrm{K}/T_{S}^{*})\right] .
\label{eq:delta-C}
\end{equation}

\begin{figure}[t]
\resizebox{\columnwidth}{!}{
\includegraphics{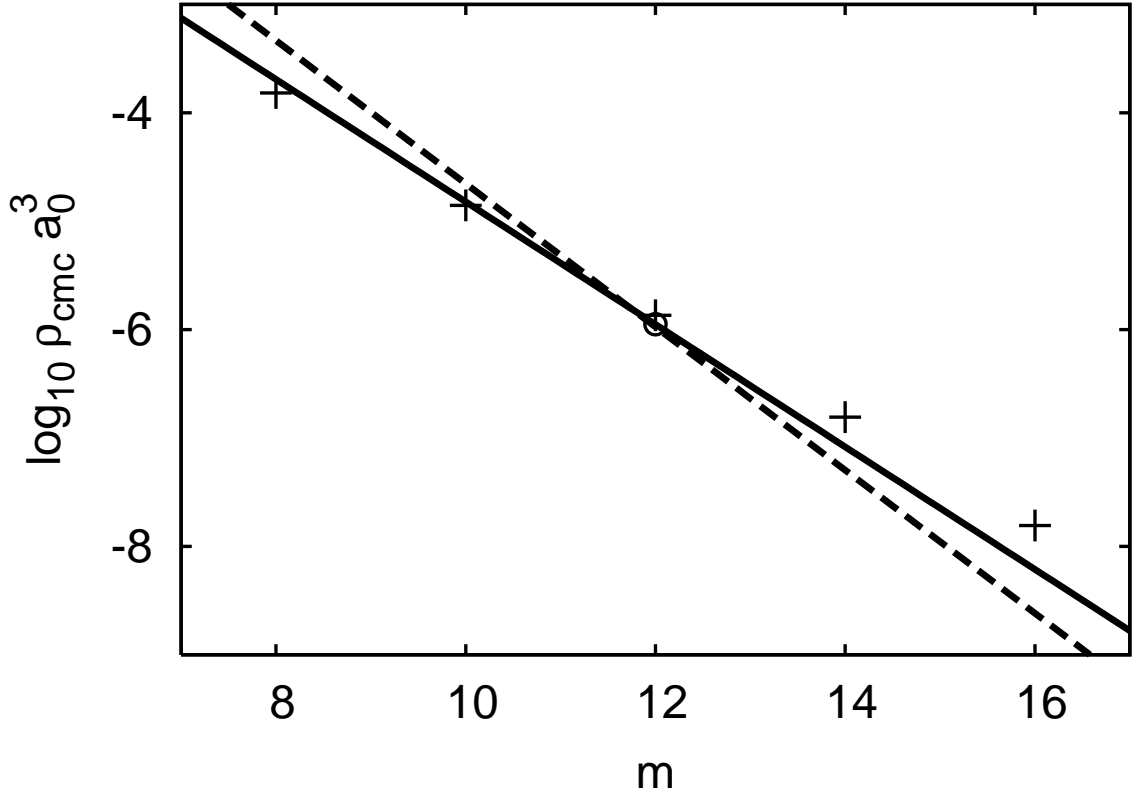}
}
\caption{\label{fig:mdependence}
Critical micelle concentration of $\mathrm{C}_m\mathrm{E}_6$ as a function of
the length of the hydrophobic tail at room temperature. Crosses and
circle are experimental data from Refs. \onlinecite{cmcbook} and
\onlinecite{Chen98}, respectively. 
A reference length $a_0 \approx  3 \, \mathrm{\AA}$ was chosen to
transform the reported values into dimensionless units. The curves show
the results of eq~\ref{eq: result} using the transfer free energy
$\DeltaMu$ (solid line) and the bulk approximation $\DeltaMu_0$
(dashed line). The length $a$ is chosen such that the experimental value at
$m=12$ from \cite{Chen98} is recovered.  Solid line: $a / a_0 = 1$,
dashed line: $a/a_0 = 1.3$, or $a/ a_0 = 1.6$ if the correction
$\Deltagamma$ to the surface tension is neglected.
}
\end{figure}

Equation \ref{eq: result} together with the dependence of transfer free
energies on alkyl chain predicts that $\cmc$ varies exponentially with
$m$. When plotted on a logarithmic scale, it should have the slope $-
\partial \DeltaMu /\partial m$. Figure \ref{fig:mdependence} compares
this prediction with experimental data. The measured data is indeed
linear over a range of concentrations spanning several orders of
magnitude. The slope of the experimental data is approximately -0.7
kcal/mol, which is close to the value of -0.9 kcal/mol inferred by
neglecting $\Delta \DeltaMu (m)$, and even closer to the value of
-0.77 kcal/mol inferred by accounting for this correction to transfer
free energies between entirely bulk phases.

\begin{figure}[t]
\resizebox{\columnwidth}{!}{
\includegraphics{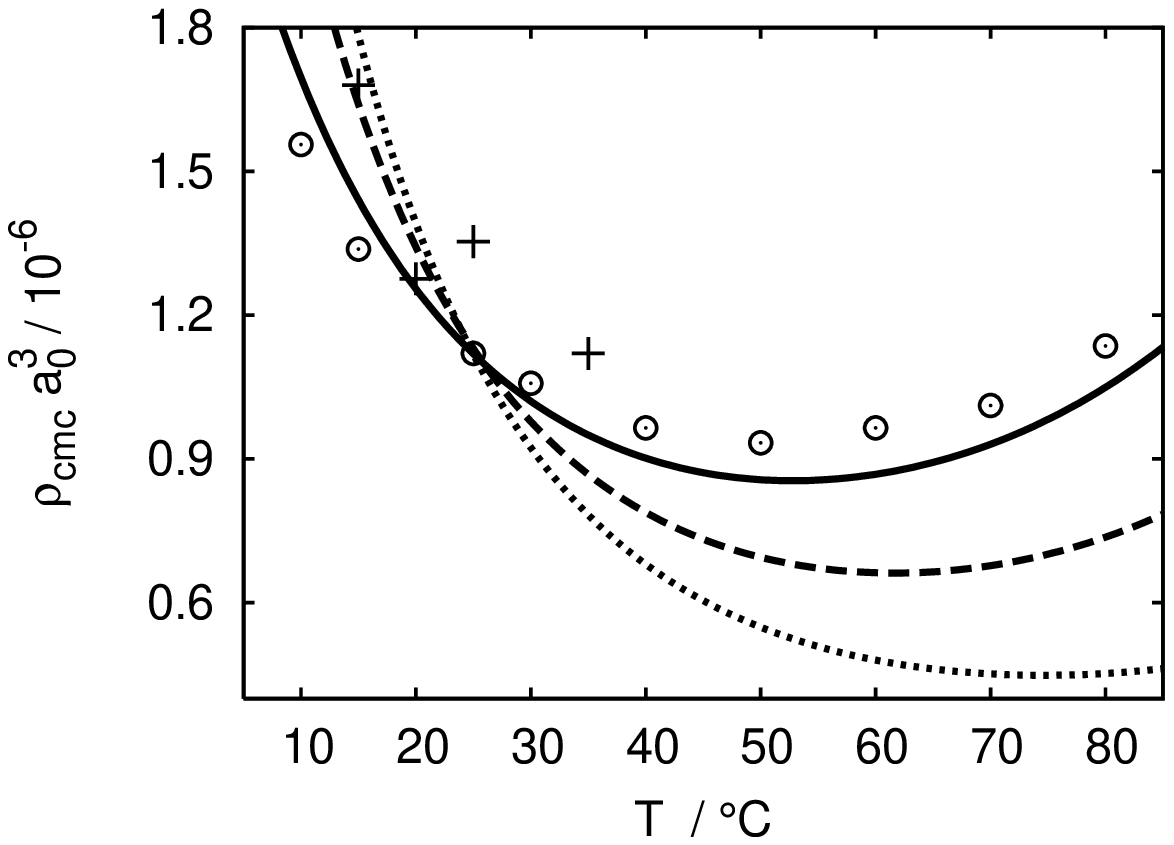}
}
\caption{\label{fig:Tdependence}
Temperature dependence of the critical micelle concentration of
$\mathrm{C}_{12}\mathrm{E}_6$. The meanings of the symbols are the same
as in Fig. \ref{fig:mdependence}. The solid line is the prediction of
eq~\ref{eq: result}. Disregarding the correction $\Delta \DeltaMu$
yields the dotted line, and additionally neglecting $\Deltagamma$
results in the
dashed line.
}
\end{figure}

Figure \ref{fig:Tdependence} concerns the temperature dependence of the
critical micelle concentration. The experimental data for $\mathrm{C}_{12}%
\mathrm{E}_{6}$ show a minimum in the critical micelle concentration near 50
Celsius. A similar minimum is found from eq \ref{eq: result}. The excellent
fit to experimental data is obtained by choosing $a$ so as to match the
experimental data at $T=25 \, ^o \mathrm{C}$. This method of choice
gives $a=3 \mathrm{\AA}$, which
is not an unreasonable value for the microscopic length. Neglecting the
corrections $\Deltagamma $ and $\Delta \DeltaMu $ lead to similar
lengths, and similar though inferior fits to the data over the observed
temperature range.

Since $\beta \gamma a^{2}$ is of order 1, and $\delta /a$ is of order 10, we
see that this expression predicts aggregation numbers of order 100. This
cluster size is consistent with our assumption of large $n$.

\section{Discussion}

Elements of the theory we have presented here for micelle assembly can be
found in earlier works. For instance, the surface energy and entropy
terms, $\DeltaG_{1}$ and $\DeltaG_{3}$, have been considered in
Ref.~\onlinecite{Woo96}. As a result, that paper obtains the same scaling of
aggregate size as given by eq~\ref{eq:n-star}, namely, $\nstar
\propto \beta \gamma \delta ^{2}$. By neglecting $\DeltaG_{2}$,
however, Ref.~\onlinecite{Woo96} errors in its treatment of the temperature
dependence of $\cmc$.

Detailed consideration of the transfer free energy contribution $\DeltaG_{2}
$ and its linearity with $m$ is found in Tanford's monograph~\cite{Tanford}.
That work does not consider $\DeltaG_1$ and $\DeltaG_3$ which
compete with $\DeltaG_2$. Both $\DeltaG_{1}$ and $\DeltaG_{2}$
contribute significantly to the temperature dependence of $\cmc$.

For the neutral surfactants we have considered, the net hydrophobic driving
force, $\DeltaG_{1}+\DeltaG_{2}$, is balanced by the super-extensive
entropy term, $\DeltaG_{3}$. It is remarkable how well $\cmc$ is
described over a wide range of conditions with only these three
terms. For ionic surfactants, an additional super-extensive term will
limit the micelle size. It is an electrostatic contribution that
opposes the clustering of like charge. 
For low concentrations of counterions, its scaling with $n$ is
identical to the entropic term, growing as $n^{5/3}$ for the case of
spherical micelles. The size of this term will depend upon
ionic strength in a fashion not yet determined.

\begin{acknowledgments}
This research has been supported by the US National Science
Foundation. Our interest in the topic of micelle formation was
rekindled by communications with Todd P. Silverstein.
\end{acknowledgments}

\section*{APPENDIX}

\label{sec:appendix}Here we consider further the law of mass action,
considering the water-surfactant mixture as an ideal solution of surfactants
and surfactant aggregates. The free energy density is~\cite{Safran}
\begin{equation}
\beta F=\sum_{n}\left[ \rho _{n}\left( \ln \rho _{n}a^{3}-1\right) +\rho
_{n}\beta f_{n}-\beta \muS n\rho _{n}\right] .
\label{eq:freeenergyfunctional}
\end{equation}
The first term accounts for the translational entropies of the aggregates. $%
f_{n}$ is the internal free energy of an $n$-mer, and $\muS$ is the surfactant
chemical potential that controls the total surfactant density $\rho
=\sum_{n}n\rho _{n}$.

The equilibrium partitioning of densities is that which minimizes eq \ref
{eq:freeenergyfunctional}:
\begin{equation}
\rho _{n}a^{3}=\exp \left( -\beta f_{n}+n\beta \muS \right)
\end{equation}
from which the law of mass action, eq \ref{eq:MassAction}, follows.

To the extent that micelles are monodisperse, $\rho _{n}\approx \rho
_{1}\delta _{n,1}+\rho _{n^{*}}\delta _{n,n^{*}}$. Here, $\delta
_{n,n^{*}}$ refers to the Kronecker delta. Substituting this
expression into eq \ref{eq:freeenergyfunctional} gives
\begin{equation}
\beta F\approx \rho _{1}\left( \ln \rho _{1}a^{3}-1\right) +\frac{\rho -\rho
_{1}}{n^{*}}\left[ \ln \left( \frac{\rho -\rho _{1}}{n^{*}}a^{3}\right)
-1\right] +\beta \rho _{1}f_{1}+\beta \left( \frac{\rho -\rho _{1}}{n^{*}}
\right) f_{n^*}- \beta \muS \rho.
\end{equation}
In this approximation, the equilibrium partitioning of aggregated and
unaggregated surfactants is obtained by minimizing this expression with
respect to $\rho _{1}$ and $n^{*}.$ That minimization gives again eq~\ref
{eq:MassAction}, and also
\begin{equation}
0=\partial \beta F/\partial n^{*}=\left[ \frac{\partial \beta \DeltaG
    / n}{
\partial n}-\frac{1}{n}\left( \ln \rho _{1}a^{3}-\beta \DeltaG / n\right)
\right] _{n=n*} .
\label{eq:dF/dn}
\end{equation}

The critical micelle concentration is identified as the lowest
surfactant density at which a measurable fraction of surfactants, $x$,
aggregate to form micelles. As such, $n\rho _{n}=x\rho _{1},$ and the
law of mass action is $\exp \left( -\beta \DeltaG\right) =\left( x\rho _{1}a^{3}/n\right) /(\rho _{1}a^{3})^{n}$. Therefore,
\begin{equation}
\beta \DeltaG / n=(1-1/n)\ln \rho _{1}a^{3}+\left( 1/n\right) \ln \left(
n/x\right) ,  \label{eq:deltaG_over_n}
\end{equation}
and $\cmc = \rho_1 \left[ 1 + \mathcal{O}\left(
  x\right)\right]$. Equation \ref{eq:cmcgeneral} with the condition
  that $\nstar$ minimizes $\DeltaG / n$ thus follows from
  eqs~\ref{eq:dF/dn} and \ref{eq:deltaG_over_n} when $n$ is large.

An alternative definition of the critical micelle concentration is the lowest density $\rho _{1}$ at which
eq \ref{eq:dF/dn} has a non-trivial solution for $n^{*}.$ This convention is used
in Ref.~\onlinecite{Woo96}. The free energy develops a local minimum for $n>1$ at
the concentration $\rho a^{3}=\exp \left[ c^{\prime }\left( \beta \gammaOW
a^{2}\right) ^{2/3}-\beta \DeltaMu \right] $, where $c^{\prime } =
(4320/49)^{1/3} \approx 4.45$, and the location of the minimum is at $n^{*}=\beta g/(5h)$.
Comparison with eqs \ref{eq: result} and \ref{eq:n-star},
respectively, shows that this convention is in close accord with the
approach we have taken.

% Here comes the list of references:

\newpage

\begin{thebibliography}{99}

\bibitem{LCW}
Lum, K.; Chandler, D.; Weeks, J.~D. {\em J. Phys. Chem. B} {\bf 1999}, {\em 103}, 4570--4577.

\bibitem{DC}
Chandler, D. {\em Nature} {\bf 2002}, {\em 417}, 491.

\bibitem{Woo96}  
Woo, H.-J.; Carraro, C.; Chandler, D. {\em Faraday Discuss.} {\bf 1996}, {\em 104}, 183--191.

\bibitem{DeGennes}
de Gennes, P.~G. {\em J. Physique Lett. (Paris)} {\bf 1979}, {\em 40}, L69--L72.

\bibitem{Stillinger}
Stillinger, F.~H. {\em J. Chem. Phys.} {\bf 1983}, {\em 78}, 4654--4661.

\bibitem{Wu92}
Wu, D.; Chandler, D.; Smit, B. {\em J. Phys. Chem.} {\bf 1992}, {\em 96}, 4077--4083.

\bibitem{greenbook}
Chandler, D. {\em Introduction to Modern Statistical Mechanics}; Oxford University Press: New York, 1987.

\bibitem{Huang-DC02}
Huang, D.~M.; Chandler, D. {\em J. Phys. Chem. B} {\bf 2002}, {\em 106}, 2047--2053.

\bibitem{Tanford}  
Tanford, C {\em The Hydrophobic Effect: Formation of Micelles \& Biological Membranes}; Wiley Publishing: New York, 1980.

\bibitem{Barrat-Hansen}
Barrat, J.-L.; Hansen, J.-P. {\em Basic Concepts for Simple and Complex Liquids}; Cambridge University Press: Cambridge, 2003.

\bibitem{oilwatersurfacetension}  
Girifalco, L.~A.; Good, R.~J. {\em J. Phys. Chem.} {\bf 1957}, {\em 61}, 904--909.

\bibitem{CRC} {\em CRC Handbook of Chemistry and Physics 83rd ed.}; CRC Press: Boca Raton, 2003.

\bibitem{McAuliffe66}  
McAuliffe, C. {\em J. Phys. Chem.} {\bf 1966}, {\em 70}, 1267--1275.

\bibitem{McAuliffe69}
McAuliffe, C. {\em Science} {\bf 1969}, {\em 163}, 478--479.

\bibitem{Murphy90}  
Murphy, K.~P.; Privalov, P.~L.; Gill, S.~J. {\em Science} {\bf 1990}, {\em 247}, 559--561.

\bibitem{Murphy94}  
Murphy, K.~P. {\em Biophys. Chem.} {\bf 1994}, {\em 51}, 311--326.

\bibitem{Baldwin86}  
Baldwin, R.~L. {\em Proc. Natl. Acad. Sci. USA} {\bf 1986}, {\em 83}, 8069--8072.

\bibitem{Safran}  
Safran, S.~A. {\em Statistical Mechanics of Surfaces, Interfaces, and Membranes}; Addison-Wesley: Reading, 1994.

\bibitem{cmcbook}  
Mukerjee, P.; Mysels, K.~J. {\em Critical Micelle Concentrations of Aqueous Surfactant Systems}; NSRDS-NBS 36, National Bureau of Standards: Washington, DC, 1971.

\bibitem{Chen98}
Chen, L.-J.; Lin, S.-Y.; Huang, C.-C.; Chen, E.-M. {\em Coll. Surf. A} {\bf 1998}, {\em 135}, 175--181.

\end{thebibliography}
\end{document}